\documentclass[reprint,aip,superscriptaddress]{revtex4-1}
\usepackage{amsmath}
\usepackage{graphicx,color}
\usepackage{newtxmath,newtxtext}

\renewcommand{\theequation}{\arabic{section}.\arabic{equation}}
\newcommand{\e}{\mathrm{e}}
\newcommand{\g}{\mathrm{g}}
\renewcommand{\d}{\mathrm{d}}

\begin{document}
\title{Insights into photosynthetic energy transfer gained from free-energy structure: 
Coherent transport, incoherent hopping, and vibrational assistance revisited}

\author{Akihito Ishizaki}
\affiliation{Institute for Molecular Science, National Institutes of Natural Sciences, Okazaki 444-8585, Japan}
\affiliation{School of Physical Sciences, Graduate University for Advanced Studies, Okazaki 444-8585, Japan}

\author{Graham R. Fleming}
\affiliation{Department of Chemistry, University of California, Berkeley, CA 94720, USA}
\affiliation{Molecular Biophysics and Integrated Bioimaging Division, Lawrence Berkeley National Laboratory, Berkeley, CA 94720, USA}
\affiliation{Kavli Energy NanoSciences Institute at Berkeley, Berkeley, CA 94720, USA}
\date{\today}

\begin{abstract}
Giant strides in ultrashort laser pulse technology have enabled real-time observation of dynamical processes in complex molecular systems. Specifically, the discovery of oscillatory transients in the two-dimensional electronic spectra of photosynthetic systems stimulated a number of theoretical investigations exploring possible physical mechanisms of the remarkable quantum efficiency of light harvesting processes. However, the  theories employed have reached a high degree of sophistication and have become complex, making it difficult to gain insights into microscopic processes and biologically significant questions. In this work, we revisit the elementary aspects of environment-induced fluctuations in the involved electronic energies and present a simple way to understand energy flow with the intuitive picture of relaxation in a funnel-type free-energy landscape. The presented free-energy description of energy transfer reveals that typical photosynthetic systems operate in an almost barrierless regime. The approach also provides insights into the distinction between coherent and incoherent energy transfer and criteria by which the necessity of the vibrational assistance is considered.
\end{abstract}
\maketitle
\section{Introduction}
\setcounter{equation}{0}

The development of new techniques of ultrafast spectroscopy have enabled real-time observation of dynamical processes in complex chemical, biological, and material systems. In the last decade, Fleming and coworkers have applied third-order nonlinear spectroscopic techniques such as two-dimensional (2D) Fourier-transformed photon echo spectroscopy to explore photosynthetic light harvesting processes \cite{Brixner:2005wu,Cho:2005bm} and revealed the existence of long-lasting oscillatory transients in 2D spectra. \cite{Engel:2007hb, Lee:2007hq,Calhoun:2009bn}
Although earlier work found coherent beats with the use of pump-probe techniques, \cite{Savikhin:1997jp} the newer experiments stimulated a rapid increase in the number of experimental and theoretical investigations to explore possible roles that quantum effects may play in the remarkable quantum efficiency of light harvesting processes in natural and artificial systems. 
\cite{OlayaCastro:2008jz, Mohseni:2008gp, Plenio:2008ff, Rebentrost:2009hu, Ishizaki:2009ky, Nazir:2009fla, Caruso:2009ib, Panitchayangkoon:2010fw, Collini:2010fy, Caruso:2010fx, Sarovar:2010hs, Ishizaki:2010ft, Hoyer:2010fl, Fassioli:2010cf, Abramavicius:2010et, Panitchayangkoon:2011cs, Scholak:2011dy, SchlauCohen:2012dn,	Lewis:2012he, Westenhoff:2012fi, Dawlaty:2012fs, Chang:2012il, Hoyer:2012fg, 
Shabani:2012ir, Wu:2012ch, Dijkstra:2012ita, Hildner:2013by, Chin:2013ia, Gelzinis:2013ch, Rivera:2013kb, OReilly:2014it, De:2014if, Dijkstra:2015fb, Scholes:2017gw, Knee:2017jd, Dutta:2017jo, Sakamoto:2017kb, Thyrhaug:2018ei, Saito:2019cv, Kim:2019io, Munoz:2020cd, Cao:2020ba, Chan:2013il, Gelinas:2014gu, Lee:2015fi, Bredas:2016cq, Kato:2018jw}

Initially, the beats were interpreted as signatures of quantum superpositions between delocalized energy eigenstates (electronic excitons) of excited pigments, and it was argued that the unexpectedly slow dephasing could enhance the efficiency of electronic energy transfer (EET). However, it has become clear that the experimentally observed oscillations have dephasing time that persist for much longer than the theoretically predicted electronic coherence lifetime, particularly at cryogenic temperatures.\cite{Panitchayangkoon:2010fw, Ishizaki:2009ky}
Hence, the possibility of vibrational contributions was addressed.\cite{Nemeth:2010iz,Mancal:2010dj,Butkus:2012hn,YuenZhou:2012hu,CaycedoSoler:2012ib,Christensson:2012gp,Tiwari:2013dt}
A plausible explanation for the moderately long-lived spectral beats was that a quantum mixture between the electronic states and the Franck--Condon active vibrational states serves to create vibronic exciton states and  produce oscillations that exhibit picosecond dephasing times, \cite{Christensson:2012gp} while very long-lived beats likely arise from ground state vibrational wave-packets.\cite{Tiwari:2013dt} 
Furthermore, oscillatory transients in the 2D electronic spectra of the photosystem II reaction center were observed, \cite{Romero:2014jm,Fuller:2014iz} suggesting that the electronic-vibrational resonance might represent an important design principle for enabling charge separation with high quantum efficiency in oxygenic photosynthesis. 
However, it was questionable whether such electronic-vibrational mixtures could be robust and could play a role under the influence of protein-induced fluctuations at physiological temperatures. By using quantum dynamics calculations, it was demonstrated that such electronic-vibrational quantum mixtures do not necessarily play a significant role in the energy transfer and charge separation dynamics, despite contributing to the enhancement of long-lived beating in 2D electronic spectra. \cite{Fujihashi:2015kz, Monahan:2015gu, Fujihashi:2016kk, Fujihashi:2018hb}
Given the variety of pigment types and range of energy gaps in natural light harvesting systems, it is not possible to make completely general statements regarding the roles of intramolecular vibrations upon photosynthetic energy/charge transfer.
However, the results of Refs.~\onlinecite{Fujihashi:2015kz, Monahan:2015gu, Fujihashi:2016kk,Fujihashi:2018hb} suggest the need for further examination on the relevance of information provided by the oscillatory behaviors in the 2D data on the studied systems and dynamics.

As stated above, quantum dynamics calculations have helped us elucidate the nature of the experimentally observed signals. However, the theories employed have reached a high degree of sophistication and have become correspondingly complex, \cite{Jang:2008ef, Ishizaki:2009jg, Prior:2010gd, Huo:2010cu, Tao:2010gq, Kelly:2011dj, Nalbach:2011cr, Berkelbach:2012dl,Kelly:2013he,Banchi:2013eo,Jang:2014wd,Banchi:2013eo,HwangFu:2015dx,Tamascelli:2019jm}  making it difficult to draw broadly applicable conclusions about biologically significant questions regarding the physical origin of the remarkable speed and efficiency of photosynthetic light harvesting.
For example, 
(1) What is the key distinction between coherent and incoherent energy transfer? 
Does the absence of observable beats necessarily imply that the energy transfer is incoherent hopping? 
(2) What do the answers to these questions imply regarding the physical origin of the remarkable speed and efficiency of photosynthetic light harvesting?
(3) Theoretical approaches generally assume ultrafast initial excitation, meaning that the ensemble begins its evolution in phase. 
What is the difference, if any, between ultrafast laser excitation and excitation by sunlight?

In this paper, we try to describe a way of understanding and visualizing the energy flow in pigment-protein complexes (PPCs) that connects the somewhat complex quantum dynamical theories to the intuitive picture of relaxation in a funnel-type free-energy landscape, which might be rough or smooth (i.e., with or without significant energy barriers for the consecutive steps). The model also relates such an energy flow to the classical Marcus picture while properly accounting for both delocalized states and independent environmental fluctuations on the individual sites. 
The free-energy description of ultrafast energy transfer reveals that photosynthetic systems operate in a barrierless regime when absorption energy differences among pigments are $\lesssim 200\,\mathrm{cm^{-1}}$. This situation is found, for example, in the Fenna-Matthews-Olson (FMO) complex. \cite{Brixner:2005wu,Cho:2005bm,Adolphs:2006ey}
This makes the initial condition rather unimportant, suggesting that the method of excitation will not play a significant role in determining the microscopic dynamics. The free energy approach also provides insight into the importance of the vibrational contributions and the distinction between coherent and incoherent energy transfer, showing that the absence of observable beats in the spectroscopy does not necessarily imply that the energy transfer occurs by incoherent hopping.

\section{Theoretical background}
\setcounter{equation}{0}

\subsection{Hamiltonian of a pigment-protein complex}

To describe EET, we restrict the electronic spectra of the $m$-th pigment in a PPC to the ground state, $\lvert \varphi_{m {\g}} \rangle$, and the first excited state, $\lvert\varphi_{m \e}\rangle$, although higher excited states may result in nonlinear spectroscopic signals. Thus, the Hamiltonian of a PPC comprising $N$ pigments is expressed as
$
	\hat{H}_{\rm PPC}
	=
	\sum_{m=1}^N
	\sum_{a= \g,\e} 
	\hat{H}_{ma}(x_m) \lvert \varphi_{ma} \rangle \langle \varphi_{ma} \rvert
	+
	\sum_{m=1}^N \sum_{n=1}^N
	J_{mn} 
	\lvert \varphi_{m \e} \rangle \langle \varphi_{m \g} \rvert
	\otimes
	\lvert \varphi_{n \g} \rangle \langle \varphi_{n \e} \rvert
$.
Here,  $\hat{H}_{ma}(x_m)$  represents the diabatic Hamiltonian for the environmental and intramolecular vibrational degrees of freedom (DoFs), $x_m$, when the system is in the $\lvert \varphi_{ma} \rangle$ state for $a= \g$ and $a= \e$. The electronic coupling between the pigments, $J_{mn}$, may also be modulated by the environmental and nuclear DoFs. In the following, however, it is assumed that the nuclear dependence of  $J_{mn}$ is vanishingly small and the Condon-like approximation is employed as usual. The Franck--Condon transition energy of the $m$-th pigment is obtained as
\begin{align}
	E_{m}^{\rm abs}
	=
	\langle
		\hat{H}_{m \e} - \hat{H}_{m {\g}}
	\rangle_{m {\g}},
\end{align}
where the canonical average has been introduced, $\langle \dots \rangle_{ma} = \mathrm{tr}(\dots \rho_{ma}^{\rm eq})$  with the environmental equilibrium state for the $\lvert \varphi_{ma} \rangle$ state, $\hat\rho_{ma}^{\rm eq} = \e^{-\beta\hat{H}_{ma}}/\mathrm{tr}\,\e^{-\beta\hat{H}_{ma}}$. Here,  $\beta$ is the inverse temperature, $1/k_{\rm B}T$. The electronic energy of each diabatic state experiences fluctuations caused by the environmental and nuclear dynamics; these dynamics are described by the collective energy gap coordinate, such that
\begin{align}
	\hat{X}_m
	=
	\hat{H}_{m \e} - \hat{H}_{m {\g}} - E_m^{\rm abs}.
	\label{eq:collective-energy-gap}
\end{align}
By definition, the mean values of the coordinate with respect to the electronic ground and excited states are given by
\begin{subequations}
\label{mean-value}
\begin{align}
	\mu_{m {\g}} &= \langle \hat{X}_m \rangle_{m {\g}} = 0,
	\label{g-mean-value-1}
	\\
	\mu_{m \e} &= \langle \hat{X}_m \rangle_{m \e} = E_m^{\rm em} - E_m^{\rm abs},
	\label{e-mean-value-2}
\end{align}
\end{subequations}
respectively, where the emission energy has been introduced, 
\begin{align}
	E_m^{\rm em} = \langle \hat{H}_{m \e} - \hat{H}_{m {\g}} \rangle_{m {\e}}.
\end{align}
For the sake of simplicity, the contribution of intramolecular vibrational modes are not considered at the current stage. The vibrational contribution will be discussed in Sec.~\ref{sec:vib-contribution}.

\subsection{Statistics of fluctuations in electronic energy}

Let the probability distribution function for the classical collective energy gap coordinate $X_m$  be $P_{ma}(X_m)$  when the system is in the $\lvert\varphi_{ma}\rangle$ state. 
The corresponding free energy is given by $ F_{ma}(X_m) = -k_{\rm B}T \ln P_{ma}(X_m) + \mathrm{const}$. In this work, it is assumed that the environmentally induced fluctuations can be described as a Gaussian process.\cite{Kubo:1985bs} Under the assumption that  $X_m$ is described as a Gaussian random variable, the probability distribution function $P_{ma}(X_m)$ is expressed as a Gaussian function,
\begin{align}
	P_{ma}(X_m)\propto \exp\left[-\frac{1}{2\sigma_{ma}^2}(X_m - \mu_{ma})^2\right],
\end{align} 
where $\sigma_{ma}^2$ denotes the variance of $X_m$ with respect to the equilibrium state associated with the $\lvert \varphi_{ma} \rangle$ state. Hence, the free energy with respect to $X_m$ is given as a quadratic function,
\begin{align}
	F_{ma}(X_m)
	=
	\frac{k_{\rm B}T}{2\sigma_{ma}^2}(X_m-\mu_{ma})^2 + \mathrm{const.},
\end{align}
and the environmental reorganization energy associated with the optical transition to the $\lvert \varphi_{ma}\rangle$ state, $E_{ma}^{\rm R} = \lvert F_{ma}(\mu_{m {\g}}) - F_{ma}(\mu_{m {\e}}) \rvert$, is obtained as
\begin{align}
	E_{ma}^{\rm R}
	&=
	\frac{k_{\rm B}T}{2\sigma_{ma}^2}(\mu_{m {\e}} - \mu_{m {\g}})^2.
	\label{reorganization-energy-1}
\end{align}
As described in Appendix~\ref{sec:linear-repsonse-static}, the relation $\mu_{m\e} = -\beta\sigma_{m \g}^2$ is valid under the Gaussian assumption, and the variances and hence the reorganization energies  are independent of the electronic states, namely $\sigma_{m {\g}}^2 =\sigma_{m {\e}}^2 = \sigma_m^2$ and $E_{m {\g}}^{\rm R}=E_{m {\e}}^{\rm R}=E_m^{\rm R}$. Hence, eqs~\eqref{mean-value} and \eqref{reorganization-energy-1} lead to 
\begin{align}
	E_m^{\rm abs} - E_m^{\rm em}= 2E_m^{\rm R},
\end{align}
and a simple relation among the variance of the fluctuations $\sigma_m^2$, temperature $T$, and the reorganization energy $E^{\rm R}$ is obtained:
\begin{align}
	\sigma_{m}^2 
	=
	k_{\rm B}T\cdot 2E_m^{\rm R}.
	\label{eq:fluctuation-reorganization}
\end{align}
Consequently, the expressions of the free energies for the electronic ground and excited states are obtained as \cite{Marchi:1993gu,Chandler:1998wa}
\begin{subequations}
\label{site-free-energy}
\begin{align}
	F_{m {\g}}(X_m) &= \frac{1}{4E_m^{\rm R}}X_m^2,
	\label{site-g-free-energy}
	\\
	F_{m {\e}}(X_m) &= E_m^{\rm abs}  - E_m^{\rm R} + \frac{1}{4E_m^{\rm R}}(X_m + 2E_m^{\rm R}).
	\label{eq:site-e-free-energy}
\end{align}
\end{subequations}
These expressions are consistent with the environmental dynamics in which the electronic and environmental states relax from the equilibrium configuration with respect to the $\lvert \varphi_{m {\g}} \rangle$ state and to the actual equilibrium configuration in the $\lvert \varphi_{m {\e}} \rangle$ state after the vertical Franck--Condon excitation, as is formulated in Appendix~\ref{sec:linear-repsonse-dynamic}.

\section{Free energy for photosynthetic energy transfer}
\setcounter{equation}{0}

\subsection{Free energy of activation required for EET to proceed}

There is no experimental evidence of nonadiabatic transitions and radiative/nonradiative decays between $\lvert \varphi_{m {\e}} \rangle$ and $\lvert \varphi_{m {\g}} \rangle$ in light-harvesting pigment-protein complexes on the picosecond timescales, and hence, we organize the product states in order of elementary excitation number. The overall ground state with zero excitation is $\lvert 0 \rangle = \prod_{m=1}^N \lvert \varphi_{m {\g}} \rangle$, whereas the presence of a single excitation at the $m$-th pigment is expressed as $\lvert m \rangle = \lvert \varphi_{m {\e}} \rangle \prod_{k(\ne m)} \lvert \varphi_{kg} \rangle$. The corresponding expansion of the complete PPC Hamiltonian yields $\hat{H}_{\rm PPC} = \hat{H}_{\rm PPC}^{(0)} + \hat{H}_{\rm PPC}^{(1)} + \dots$, where $\hat{H}_{\rm PPC}^{(n)}$  describes $n$-excitation manifold comprising $n$ elementary excitations. The Hamiltonian of the zero-excitation manifold is 
\begin{align}
	\hat{H}_{\rm PPC}^{(0)} 
	= 
	\hat{H}^{(0)}\lvert 0 \rangle\langle 0 \rvert
	\label{0em-hamiltonian}
\end{align}	
with $\hat{H}^{(0)}=\sum_{m=1}^N \hat{H}_{m {\g}}$.
The Hamiltonian of the one-excitation manifold takes the form,
\begin{align}
	\hat{H}_{\rm PPC}^{(1)}
	=
	\sum_{m=1}^N \hat{H}_m \lvert m \rangle \langle m \rvert
	+
	\sum_{m,n} J_{mn} \lvert m \rangle\langle n \rvert,
	\label{1em-hamiltonian}
\end{align}	
where $\hat{H}_m$ has been introduced as 
\begin{align}
	\hat{H}_m
	&=
	\hat{H}_{m {\e}} + \textstyle\sum_{k(\ne m)}\hat{H}_{kg} 
	\notag\\
	&=
	E_m^{\rm abs} + \hat{X}_m + \hat{H}^{(0)}.
	\label{m-site-hamiltonian}
\end{align}
The intensity of sunlight is weak and, thus, the single-excitation manifold is of primary importance under physiological conditions, although nonlinear spectroscopic techniques such as 2D electronic spectroscopy can populate higher excitation manifolds.

\begin{figure}
	\includegraphics{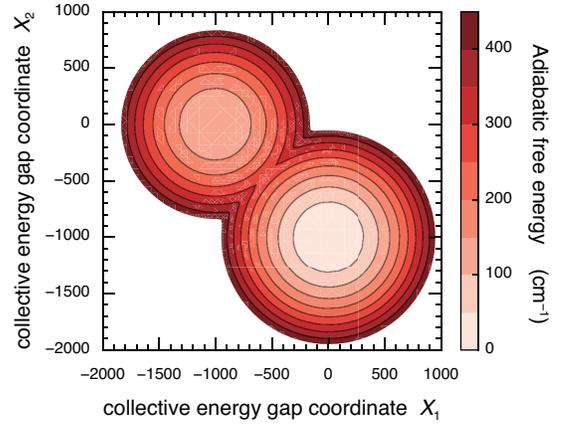}
	\caption{Contour plot of the adiabatic free-energy surface of a dimer as a function of the two collective energy gap coordinates, $X_1$ and $X_2$, which describe fluctuations in the electronic energies.
	The parameters are chosen to be $E_1^{\rm abs}-E_2^{\rm abs}=100\,\mathrm{cm^{-1}}$, $J_{12} = -10\,\mathrm{cm^{-1}}$, and $E_1^{\rm R}=E_2^{\rm R}=500\,\mathrm{cm^{-1}}$. Contour lines are drawn at $50\,\mathrm{cm^{-1}}$ intervals. 
In the case of large reorganization energy, the adiabatic free-energy surface $F_-(X_1,X_2)$ typically possesses two local minima corresponding to the environmental equilibria associated with states $\lvert 1 \rangle = \lvert \varphi_{1 \e} \rangle\lvert \varphi_{2 \g} \rangle$ and $\lvert 2 \rangle = \lvert \varphi_{1 \g} \rangle\lvert \varphi_{2 \e} \rangle$. The point of origin corresponds to the Franck--Condon state. Incoherent hopping EET taking place from one minimum to another requires overcoming the free-energy barrier via thermal activation.}
	\label{fig:1}
\end{figure}

Correspondingly to the Hamiltonians, the free energy surfaces for the zero- and one-excitation manifolds are given as
\begin{align}
	F^{(0)}({\bf X})
	&=
	\sum_{m=1}^N F_{m {\g}}(X_m),
\\
	F^{(1)}({\bf X})
	&
	=
	\sum_{m=1}^N F_m({\bf X}) \lvert m \rangle\langle m \rvert 
	+
	\sum_{m,n} J_{mn} \lvert m \rangle \langle n \rvert,
	\label{1em-free-energy}
\end{align}
where the free-energy $F_m({\bf X})$ has been introduced correspondingly to the Hamiltonian in eq~\eqref{m-site-hamiltonian} as 
\begin{align}
	F_m({\bf X})
	&=
	F_{m {\e}}(X_m) + \textstyle\sum_{k(\ne m)}F_{kg}(X_k)
	\notag\\
	&=
	E_m^{\rm abs} + X_m + F^{(0)}({\bf X}).
\end{align}
For simplicity, the study addresses a dimer comprising pigments 1 and 2 and considers the two-dimensional reaction coordinate space, ${\bf X} = (X_1, X_2)$.
In this case, eq~\eqref{1em-free-energy} is easily diagonalized, and the adiabatic free energy surfaces in the one-excitation manifold are obtained. Figure~\ref{fig:1} draws the adiabatic free energy surface as a function of the two collective energy gap coordinates, $X_1$ and $X_2$. For demonstration purpose, the parameters are chosen to be $E_1^{\rm abs}-E_2^{\rm abs}=100\,\mathrm{cm^{-1}}$, $J_{12} = -10\,\mathrm{cm^{-1}}$, and $E_1^{\rm R}=E_2^{\rm R}=500\,\mathrm{cm^{-1}}$. In the case of weak electronic coupling, the adiabatic free-energy surface $F_-(X_1,X_2)$ typically possesses two local minima corresponding to the environmental equilibria associated with states $\lvert 1 \rangle = \lvert \varphi_{1e} \rangle\lvert \varphi_{2g} \rangle$ and $\lvert 2 \rangle = \lvert \varphi_{1g} \rangle\lvert \varphi_{2e} \rangle$. The point of origin corresponds to the Franck--Condon state. The free energy of the point is higher than the barrier between the minima; therefore, delocalized excitons may be found immediately after the excitation even in the F\"orster regime, depending on the magnitude of the electronic coupling. \cite{Ishizaki:2009jg, Ishizaki:2010fx} As time increases, dissipation of reorganization energy proceeds and the excitation will fall into one of the minima and become localized, and subsequent {\it incoherent hopping EET taking place from one minimum to another requires overcoming the free-energy barrier via thermal activation.}

\begin{figure}
	\includegraphics{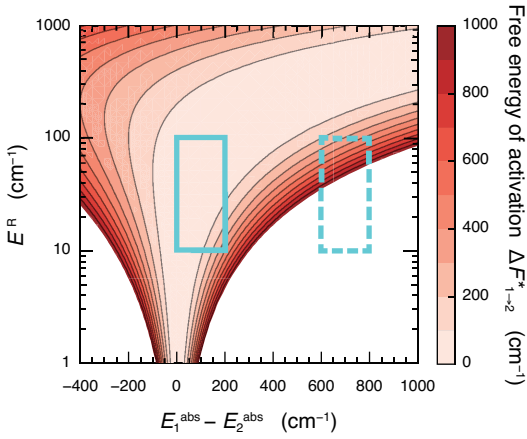}
	\caption{
	Contour plot of the free energy of activation required for the electronic energy transfer to proceed. The free energy of activation $\Delta F^\ast_{1\to 2}$ is plotted as a function of the difference between the absorption energies $E_1^{\rm abs} - E_2^{\rm abs}$ and the reorganization energy $E^{\rm R}=E^{\rm R}_1=E^{\rm R}_2$. Contour lines are drawn at $100\,\mathrm{cm^{-1}}$ intervals. 
	The box with the solid line indicates the typical values for the photosynthetic electronic energy transfer, $0 \le E_1^{\rm abs}-E_2^{\rm abs} \le 200\,\mathrm{cm^{-1}}$ and $10\,\mathrm{cm^{-1}}\le E^{\rm R} \le 100\,\mathrm{cm^{-1}}$. \cite{vanAmerongen:2000fy}
	The thermal energy is evaluated as $k_{\rm B}T \simeq 200\,\mathrm{cm^{-1}}$ at physiological temperature $T = 300\,{\rm K}$, and therefore, for energy gap $\le 200\,\mathrm{cm^{-1}}$ EET takes place in an almost activationless fashion in the parameter region corresponding to natural photosynthesis. For large energy gap an activation free energy is required (dashed box).}
	\label{fig:2}
\end{figure}

A question naturally arises concerning the height of the free energy barrier or the free energy of activation. In the absence of the electronic coupling, the intersection of the diabatic free energy surfaces, $F_1(X_1,X_2)$ and $F_2(X_1,X_2)$, is expressed as $X_1+E_1^{\rm R} = X_2 + E_2^{\rm R}$, and the coordinates of the saddle point are given by 
$X_1^\ast = {E_1^{\rm R}(E_2^{\rm abs}-E_1^{\rm abs}-2E_2^{\rm R})}/(E_1^{\rm R}+E_2^{\rm R})$, and $X_2^\ast = {E_2^{\rm R}(E_1^{\rm abs}-E_2^{\rm abs}-2E_1^{\rm R})}/(E_1^{\rm R}+E_2^{\rm R})$. Hence, the height of the free-energy barrier associated with the transfer from pigments 1 to 2 is given by a simple formula similar to the Marcus' energy gap law, \cite{Marcus:1993kx}
\begin{align}
	\Delta F^\ast_{1\to 2}
	=
	\frac{(E_1^{\rm abs}-E_2^{\rm abs}-2E_1^{\rm R})^2}{4(E_1^{\rm R}+E_2^{\rm R})}.
	\label{free-energy-barrier-1}
\end{align}
In deriving eq~\eqref{free-energy-barrier-1}, the electronic coupling $J_{mn}$ has been assumed to be vanishingly small for simplicity. The finite value of the coupling lowers the free-energy barrier by $J_{mn}$. Equation~\eqref{free-energy-barrier-1} is regarded as a multidimensional extension of Marcus theory; however, it can be recast in terms of the donor emission energy as $\Delta F^\ast_{1\to 2} = {(E_1^{\rm em}-E_2^{\rm abs})^2}/{4(E_1^{\rm R}+E_2^{\rm R})}$. This is physically consistent with the F\"orster rate formula expressed by the overlap integral of the donor-emission lineshape $F_1(\omega)$ and the acceptor-absorption lineshape $A_2(\omega)$.

Figure~\ref{fig:2} presents the free energy of activation, $\Delta F_{1\to 2}^\ast$, as a function of the reorganization energy, $E^{\rm R}=E_1^{\rm R}=E_2^{\rm R}$ and the absorption energy difference, $E_1^{\rm abs}-E_2^{\rm abs}$. For typical values of $0 \le  E_1^{\rm abs}-E_2^{\rm abs} \le 200\,\mathrm{cm^{-1}}$ and $10\,\mathrm{cm^{-1}}\le E^{\rm R} \le 100\,\mathrm{cm^{-1}}$, \cite{vanAmerongen:2000fy} the free energy of activation is very small in comparison to the thermal energy, $k_{\rm B}T\simeq 200\,\mathrm{cm^{-1}}$ at physiological temperature $T=300\,{\rm K}$. In other words, EET takes place in a practically activationless fashion. This clarifies the physical origin of ultrafast EET. 
In Figure~2 of Ref.~\onlinecite{Ishizaki:2009jg}, it is shown that the EET rate in the case of $ E_1^{\rm abs}-E_2^{\rm abs} = 100\,\mathrm{cm^{-1}}$ is maximized for values of the reorganization energy typical in natural light harvesting systems.
This optimization is consistent with the activationless nature of the free-energy surface. In natural light harvesting systems, there is a manifold of states.
Although the coordinates involved span multiple dimensions, this can be visualized as relaxation down a slightly ``bumpy'' funnel in a rather similar fashion to the Wolynes' picture of protein folding landscapes. \cite{Bryngelson:1987uu} Furthermore, the barrierless nature of the energy transfer over the wide range of parameters typical in natural light harvesting systems means that inhomogeneous broadening has rather little influence on the dynamics. Similarly, there should be a very weak temperature dependence as observed, for example, in the LH2 complex of purple bacteria. \cite{Yang:2001bw}

\subsection{Coherent versus incoherent}

In the literature, the term of ``coherent transfer'' indicates that excitation travels as a quantum mechanical wave packet keeping its phase coherence; otherwise, the term of ``incoherent transfer'' is employed. As has been already discussed above, the incoherent ``hopping'' takes place from one minimum on the free-energy surface to another by overcoming a free-energy barrier that requires thermal activation. As demonstrated in eq~\eqref{free-energy-barrier-1} and Figure~\ref{fig:2}, however, the barrier is insignificantly small in comparison to the thermal energy when energy gaps of $\le 200\,\mathrm{cm^{-1}}$ are considered, and thus the EET takes place in a nearly activationless fashion. Although this transfer process does not have coherent dynamics in the sense of the macroscopic ensemble, the transfer is still mainly driven by the electronic interaction between the pigments rather than by the thermal activation. The absence of long-lasting oscillatory transients in the ensemble average does not necessarily provide a correct insight into the microscopic nature of the EET dynamics.\cite{Ishizaki:2011cx}
Therefore, activationless EET needs to be distinguished from genuine incoherent ``hopping'' that requires thermal activation.

The current approach also has an implication regarding the importance of the initial excitation, either by pulsed coherent light, by incoherent thermal light such as sunlight, or by energy transfer. If the barrier for energy transfer is substantial, the initial condition is influenced accordingly. That is, on a bumpy free energy landscape with a free-energy well traps, the initial state would determine the course of the EET. However, if the process is barrierless, the sensitivity on the initial condition is greatly reduced.

\subsection{Vibrational contribution\label{sec:vib-contribution}}

\begin{figure}
	\includegraphics{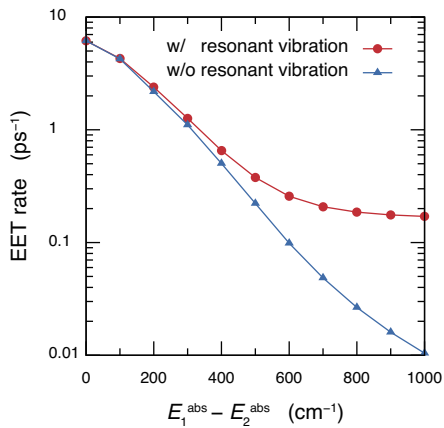}
	\caption{EET rate influenced by a resonant vibrational mode whose frequency is  $\omega_{\rm vib} = E_1^{\rm abs} - E_2^{\rm abs}$. The calculated rates are presented as a function of the absorption energy difference, $E_1^{\rm abs} - E_2^{\rm abs}$. They were obtained in the same fashion as in Ref.~\onlinecite{Fujihashi:2015kz} with the following parameters: 
	the inter-site electronic coupling, $J_{12} = 50\,\mathrm{cm^{-1}}$; the environmental reorganization energy, the reorganization time constant, and the temperature set to $\lambda_{\rm env} = 35\,\mathrm{cm^{-1}}$,  $\tau_{\rm env} = 100\,{\rm fs}$, and $T=300\,{\rm K}$, respectively; the vibrational relaxation time constant and the Huang-Rhys factor set to $\gamma_{\rm vib}^{-1}=2\,{\rm ps}$ and $S=0.025$, respectively. The Condon approximation was employed for the transition dipole moments.
	}
	\label{fig:3}
\end{figure}

The examination of the free energy of activation can also provide an insight into the necessity of a vibrational contribution to assist the EET. In the case that the free energy of activation does not exceed the thermal energy, the EET can take place easily without the assistance of vibrational modes, even though spectroscopic measurements may detect vibrational and vibronic signatures. This situation corresponds to the parameter region marked with the solid line box in Figure~\ref{fig:2}. Indeed, Ref.~\onlinecite{Fujihashi:2015kz} demonstrated that the electronic-vibrational quantum mixtures do not necessarily play a significant role in EET dynamics in the FMO complex, despite contributing to the enhancement of long-lived quantum beating in 2D electronic spectra. In other photosynthetic pigment-protein complexes, however, relatively large differences among the absorption energies are found, e.g., in 
light harvesting complex II (LHCII),\cite{Novoderezhkin:2005ey,SchlauCohen:2009bo,Bhattacharyya:2020kn,Arsenault:2020dh} 
phycoerythrin 545, 
\cite{Collini:2010fy,Kolli:2012ip,OReilly:2014it} 
phycocyanin 645. 
\cite{Dean:2016ei,Blau:2018ce,Bennett:2018hb}
Hence, the parameter region marked with the dashed line box in Figure~\ref{fig:2} was addressed, where $600\,\mathrm{cm^{-1}} \le   E_1^{\rm abs}-E_2^{\rm abs} \le 800\,\mathrm{cm^{-1}}$ and $10\,\mathrm{cm^{-1}}\le E^{\rm R} \le 100\,\mathrm{cm^{-1}}$. In this region, the free energy of activation is high compared to the thermal energy at physiological temperatures.

Some possible options for lowering the barrier with the fixed absorption energy difference are: (1) an increase in the environmental reorganization energy $E^{\rm R}$,  (2) an increase in the inter-site electronic coupling  $J_{mn}$, (3) correlated fluctuations in electronic energies,\cite{Lee:2007hq,Ishizaki:2010ft} or (4) assistance by vibrational modes. When the energy acceptor is a vibrationally excited state of pigment 2, the absorption energy difference is reduced to $E_1^{\rm abs}-(E_2^{\rm abs} + \hbar\omega_{\rm vib})$. In particular, the resonance between the vibrational frequency and the absorption energy difference leads to $E_1^{\rm abs}-(E_2^{\rm abs} + \hbar\omega_{\rm vib}) = 0$, and thus, the free-energy barrier in Figure~\ref{fig:2} is substantially lowered. 
When the Condon approximation is valid for the transition dipole moments, the inter-pigment coupling is reduced as $J_{mn} \to -J_{mn} \e^{-S/2}\sqrt{S}$, with $S$ being the Huang-Rhys factor of the vibrational mode.
Nevertheless, a substantial rate enhancement by a high-frequency vibrational mode is possible, as demonstrated in Figure~\ref{fig:3}.
This is consistent with the recent theoretical result on vibrationally assisted EET in LHCII.\cite{Bhattacharyya:2020kn}
When the non-Condon effect is more prominent, the effect results in the enhancement of the vibronic transitions, \cite{Zhang:2016in} possibly leading to the further acceleration of the vibrationally assisted EET.\cite{Arsenault:2020dh}

Figure~\ref{fig:3} presents the EET rate influenced by a resonant vibrational mode whose frequency is a function of the absorption energy difference, $E_1^{\rm abs} - E_2^{\rm abs}$. Consistently with the insight gained from the free-energy barrier in Figure~\ref{fig:2}, the extent of the vibrational assistance increases with increasing the value of $E_1^{\rm abs}-E_2^{\rm abs}$. Here, it is noted that the vibrationally assisted EET rate exhibits a plateau in the region of $E_1^{\rm abs} - E_2^{\rm abs} > 700\,\mathrm{cm^{-1}}$, indicating that the EET is promoted only by the vibrationally excited state. This can be also understood via the F\"orster rate formula. In the presence of a high-frequency vibrational mode, the absorption lineshape of pigment 2 is expressed as $A_2(\omega) = A_2^{(0)}(\omega) + A_2^{(1)}(\omega) + \dots$, where $A_2^{(v)}(\omega)$ is the lineshape associated with the optical transition to the $v$-th vibrational level in the electronically excited state and is approximately expressed as $	A_2^{(v)}(\omega) \simeq ({S^v}/{v!}) A_2^{(0)}(\omega-v\omega_{\rm vib})$. In the case of high-frequency vibrational mode, 
the overlap between $A_2^{(0)}(\omega)$ and $A_2^{(1)}(\omega)$ vanishes, and hence, the vibrationally assisted EET rate is evaluated with the overlap integral between $F_1(\omega)$ and $A_2^{(1)}(\omega)= S A_2^{(0)}(\omega-\omega_{\rm vib})$. It should be noted that the overlap integral is independent of the value of $ E_1^{\rm abs}-E_2^{\rm abs}$ under the condition of $ E_1^{\rm abs}-E_2^{\rm abs} = \omega_{\rm vib}$.

\section{Concluding remarks}

This paper presented a description of the photosynthetic energy transfer using a free-energy surface in order to unveil the physical origins of its remarkable speed and to connect the complex quantum dynamical models to the intuitive picture of energy flow in photosynthetic pigment-protein complexes. 
The presented free-energy description of ultrafast energy transfer reveals that photosynthetic systems operate in the barrierless regime when absorption energy differences among the pigments are $\lesssim 200\,\mathrm{cm^{-1}}$.
This makes the initial condition rather unimportant, suggesting that the method of excitation will not play a significant role in determining the microscopic processes. Our approach also provides insight into the necessity of the vibrational contribution and the distinction between coherent and incoherent energy transfer, showing that the absence of observable beats in the spectroscopy does not necessarily imply that the energy transfer is incoherent hopping.
It is noted that the conclusions are valid when relatively weak electronic coupling, small Huang-Rhys factors and the Condon approximation can be employed. Although these are often the appropriate regime for photosynthetic light harvesting, they are not always so. For such cases, more careful treatment should be necessary.

Although ways to electronically excite pigments will not play a significant role in determining the microscopic processes in the parameter region typical of photosynthetic light harvesting systems, it is still intriguing to elucidate how photo-excitation by natural light and the subsequent dynamics proceed with quantitative underpinnings. There are multiple spatiotemporal hierarchies related to the interaction between molecules and natural light, and thus, an issue similar to the above ``coherent vs incoherent'' is also present. 
In the timescales of direct human observation, sunlight photons continuously pump photosynthetic systems, and therefore, the excitation is sometimes considered as by an incoherent continuous wave \cite{Brumer:2012ib} or modeled as a photon bath. \cite{Mancal:2010kc,Fassioli:2012gd,Pachon:2017la} As discussed in Ref.~\onlinecite{Blankenship:2013th}, however, the sunlight flux is estimated to be about $ 10\,{\rm s^{-1}} {\rm \AA^{-2}}$ at full sunlight, and the number of photons absorbed by a chlorophyll molecule is at most $10\,{\rm s^{-1}}$. On sub-picosecond and nanometer scales, consequently, the photon density is vanishingly small and a single photosynthetic pigment will be influenced by only a single photon. \cite{Chan:2018em} In such a situation, photons may not be treated as continuous or pulsed waves in a classical manner, because the single-photon state $\lvert \psi \rangle$ leads to $\langle \psi \vert \hat{E}({\bf r}, t) \vert \psi \rangle = 0$ for the electric field operator, $\hat{E}({\bf r}, t)$. Therefore, it might be necessary to take into account the quantum mechanical nature of the photons. 
Recently, a nonclassical Hong-Ou-Mandel interference between sunlight and single photons from a semiconductor quantum dot was experimentally demonstrated.\cite{Deng:2019cu}
The theoretical investigation on pseudo-sunlight through the use of quantum entangled photons and its interaction with molecules \cite{Fujihashi:2020ep} is also an attempt along this line.
More elaborate investigations of photo-excitation by natural light and the subsequent excited-state dynamics in molecular systems are left for future studies.

\begin{acknowledgments}
The authors are grateful to Yuta Fujihashi for drafting the figures.
This work was supported by JSPS KAKENHI Grant No.~17H02946, MEXT KAKENHI Grant No.~17H06437 in Innovative Areas ``Innovations for Light-Energy Conversion,'' and MEXT Quantum Leap Flagship Program Grant No.~JPMXS0120330644.
GRF's contribution was supported by the US Department of Energy, Office of Science, Basic Energy Sciences, Division of Chemical Sciences, Geosciences, and Biosciences.
\end{acknowledgments}

\appendix
\section{Static properties of the environmental degrees of freedom}
\label{sec:linear-repsonse-static}
\renewcommand{\theequation}{\ref{sec:linear-repsonse-static}.\arabic{equation}}

The definition of the collective energy gap coordinate in eq~\eqref{eq:collective-energy-gap} is recast into 
\begin{align}
	\hat{H}_{m {\e}} 
	= 
	\hat{H}_{m {\g}} + \hat{X}_m + E_{m {\g}},
\end{align} 
and thus, $\hat{H}_{m {\g}}$ and $\hat{X}_m$ may be regarded as an unperturbed system Hamiltonian and an external field, respectively. The linear response theory \cite{Kubo:1985bs} allows to approximate the environmental equilibrium state associated with $\lvert \varphi_{m \e} \rangle$ as
\begin{align}
	\hat\rho_{m {\e}}^{\rm eq}
	=
	\hat\rho_{m {\g}}^{\rm eq}
	\left(
		1 
		- 
		\int^\beta_0 \d \lambda\,
		\e^{\lambda \hat{H}_{m {\g}}}\hat{X}_{m} \e^{-\lambda \hat{H}_{m {\g}}}
	\right).
\end{align}
Consequently, the mean value of the collective energy gap coordinate with respect to the electronic excited state $\mu_{m {\e}} = \langle \hat{X}_m \rangle_{m \e}$ is evaluated as
\begin{align}
	\mu_{m {\e}}
	=
	-\beta \langle \hat{X}_m ; \hat{X}_m \rangle_{m {\g}},
	\label{appendix:e-mean-value-2}
\end{align}
where $\langle \hat{X}_m \rangle_{m \g} = 0$ in eq~\eqref{mean-value} has been employed, and $\langle \hat{O}_1;\hat{O}_2 \rangle_{m a}$ stands for the canonical correlation \cite{Kubo:1985bs} defined by
$
	\langle \hat{O}_1 ; \hat{O}_2 \rangle_{m a} 
	= 
	\beta^{-1}\int^\beta_0 \d\lambda 
	\langle \e^{\lambda \hat{H}_{m a}} \hat{O}_1 \e^{-\lambda\hat{H}_{ma}}\hat{O}_2 \rangle_{m a}
$.
In the classical limit of $\hbar\to 0$, the canonical correlation function can be approximated as the classical correlation function, and thus eq~\eqref{appendix:e-mean-value-2} yields
\begin{align}
	\mu_{m {\e}}
	=
	-\beta \sigma_{m \g}^2.
	\label{appendix:g-valiance}
\end{align}
Furthermore, the variance of the coordinate with respect to the electronic excited state is evaluated as
\begin{align}
	\sigma_{m {\e}}^2
	=
	\langle \hat{X}_m^2 \rangle_{m {\e}}
	=
	\mathrm{tr}(\hat{X}_m^2 \hat\rho_m^{\rm eq})
	=
	\langle \hat{X}_m^2 \rangle_{m {\g}}
	=
	\sigma_{m {\g}}^2,
	\label{appendix:e-valiance}
\end{align}
where three-body correlation functions such as $\langle \hat{X}_m ; \hat{X}_m^2 \rangle_{m {\g}}$ have vanished owing to the Gaussian statistics. As a consequence, eq~\eqref{reorganization-energy-1} leads to $E_{m {\g}}^{\rm R} = E_{m {\e}}^{\rm R}$.

\section{Dynamic properties of the environmental degrees of freedom}
\label{sec:linear-repsonse-dynamic}
\renewcommand{\theequation}{\ref{sec:linear-repsonse-dynamic}.\arabic{equation}}

After the electronic excitation in accordance with the vertical Franck--Condon transition, the electronic states and their surrounding environment reorganize from the equilibrium configuration with respect to the electronic ground state to reach the actual equilibrium configuration in the excited state. 
To formulate this process, let $\hat\rho_{m {\e}}(t) $ be the density operator that describes dynamics of the environmental DoFs associated with the electronically excited state of the $m$-th pigment. 
The time-evolution after the photoexcitation at $t=0$ is described with the Hamiltonian,
\begin{align}
	\hat{H}_{m {\e}}(t) = \hat{H}_{m {\g}} + ( \hat{X}_m + E_{m {\g}} )\theta(t),
\end{align}
with $\theta(t)$ being the Heaviside step function, and hence the Liouville equation is written as
\begin{align}
	\frac{\partial}{\partial t}\hat\rho_{m {\e}}(t) 
	=
	-\frac{i}{\hbar} \left[ \hat{H}_{m {\g}}+\hat{X}_m\theta(t),\hat\rho_{m {\e}}(t) \right].
\end{align}
Similarly to Appendix \ref{sec:linear-repsonse-static}, $\hat{H}_{m {\g}}$ and $\hat{X}_m\theta(t)$ can be regarded as an unperturbed system Hamiltonian and a time-dependent external field, respectively. The linear response theory \cite{Kubo:1985bs} allows to approximate $\hat\rho_{m {\e}}(t)$ as 
\begin{align}
	\hat\rho_{m {\e}}(t)
	=
	\hat\rho^{\rm eq}_{m {\g}}
	-
	\frac{i}{\hbar}
	\int^t_{0} \d s\, 
	\hat{\mathcal{G}}_{m {\g}}(t-s)
	\left[ \hat{X}_m, \hat\rho^{\rm eq}_{m {\g}} \right],
	\label{eq:env-dynamics}
\end{align}
where $\hat{\mathcal{G}}_{m {\g}}(t)$ is the time-evolution operator in the Liouville space, $\hat{\mathcal{G}}_{m {\g}}(t)\hat{O} = \e^{-i\hat{H}_{m {\g}}t/\hbar} \hat{O} \e^{i\hat{H}_{m {\g}}t/\hbar}$ for any operator $\hat{O}$.
The environmental dynamics can be measured by using the time-dependent fluorescence Stokes shift experiment. The experiment records the nonequilibrium energy difference between the electronic ground and excited states as a function of the delay time $t$ after the photoexcitation, 
\begin{align}
	\Delta E_m(t) = \mathrm{tr} \left[ (\hat{H}_{m {\e}} - \hat{H}_{m {\g}})\hat\rho_{m {\e}}(t) \right].
\end{align}
Substituting eq~\eqref{eq:env-dynamics}, we obtain 
\begin{align}
	\Delta E_m(t) = E_m^{\rm abs} - \Psi_m(0) + \Psi_m(t), 
	\label{eq:nonequilibrium-energy-difference}
\end{align}
with $\Psi_m(t)$ being the relaxation function defined by
\begin{align}
	\Psi_m(t)
	=
	\beta\langle \tilde{X}_m(t); \tilde{X}_m(0)\rangle_{m {\g}},
	\label{eq:canonical-correlation-dynamics}
\end{align}
where the notation $\tilde{X}_m(t) = \e^{i\hat{H}_{m {\g}}t/\hbar} \hat{X}_m \e^{-i\hat{H}_{m {\g}}t/\hbar}$ has been introduced. According to eqs~\eqref{appendix:e-mean-value-2} and \eqref{mean-value}, the identity $\Psi_m(0) = - \mu_{m\e} = E_m^{\rm abs} - E_m^{\rm em}$ holds valid. Typically, the relaxation function $\Psi_m(t)$ converges to 0 in the long time limit, and therefore, $\Delta E(0) = E_m^{\rm abs}$ and $\Delta E(\infty) = E_m^{\rm em}$ are obtained.


\end{document}